\documentclass[9pt,twocolumn,twoside]{osajnl}

\journal{pr} 

\usepackage{amsmath}
\usepackage{fdsymbol}
\usepackage{multirow}
\usepackage{multicol}
\usepackage{stfloats}
\usepackage{graphicx}
\usepackage{wrapfig}
\graphicspath{{figures/}}

\setboolean{shortarticle}{false}

\title{Non-iterative complex wave-field reconstruction based on Kramers-Kronig relations}

\author[1,*]{Cheng Shen}
\author[2]{An Pan}
\author[1]{Mingshu Liang}
\author[1]{Changhuei Yang}

\affil[1]{Department of Electrical Engineering, California Institute of Technology, Pasadena, CA 91125, USA}
\affil[2]{Xi'an Institute of Optics \& Precision Mechanics (XIOPM), Chinese Academy of Sciences (CAS), Xi'an 710119, China}
\affil[*]{Corresponding author: cshen3@caltech.edu}

\begin{abstract}
A new computational imaging method to reconstruct the complex wave-field is reported. Due to the existence of zero frequency component, the measured signal by amplitude modulation of pupil has a spectrum similar to the one of off-axis hologram. The mathematical analogy between them is established in this paper. Based on this observation and analyticity of band-limited signal under any diffraction-limited system, an algorithm from Kramers-Kronig (KK) relations is utilized to recover the phase information only from the intensity patterns. From the sensing side, only two measurements are required at least. From the reconstruction algorithm side, our method is iteration-free and parameter-free, also without any assumption on sample characteristics. It owns several advantages over existing phase imaging methods and could provide a unique perspective to understand current computational imaging methods.
\end{abstract}

\setboolean{displaycopyright}{true}

\begin{document}

\maketitle
\section{Introduction}
Optical wave-field can be described as a 2-D complex function under the scalar diffraction theory \cite{goodman2005introduction}. This is a simple yet powerful model. The difficulty to obtain the whole information of complex wave-field lies in the phase detection. Due to the insufficient response rate of even state-of-the-art detectors compared to the electromagnetic frequency of light, direct phase measurement is not viable. Although an interesting work just came out using compressed ultrafast photography to image light-speed phase signals in a single shot \cite{kim2020picosecond}, its performance is still rudimentary and hardly incorporated into many imaging setups. Traditional quantitative phase measurement methods can be generally categorized as interferometric and non-interferometric. The former includes digital holography \cite{marquet2005digital,seelamantula2011exact,popescu2006diffraction,wang2011spatial,anand2018tutorial} and phase-shifting interferometry \cite{shaked2009two,gao2011parallel} and optical coherence tomography \cite{joo2005spectral,yaqoob2009improved}. Non-interferometric methods can be further divided into iterative phase (diversity) retrieval \cite{gerchberg1972practical,fienup1982phase,shen2017two}, (Fourier) ptychography \cite{rodenburg2004phase,pan2019three,zheng2013wide,pan2018subwavelength}, transport of intensity equation \cite{waller2010transport,zuo2013transport}, and quantitative differential phase contrast method \cite{mehta2009quantitative,tian2015quantitative,lu2016quantitative}. These popular phase imaging modalities have found numerous applications in varieties of fields, including cellular mechanics and biophysics \cite{park2008refractive,eldridge2019comparing}, digital pathology \cite{greenbaum2014wide,horstmeyer2015digital}, X-ray crystallography \cite{miao1998phase}, and optical metrology \cite{anand2009shape}.

In this letters, a new computational imaging method to reconstruct complex wave-field is proposed. Since it stitches different frequency bands in Fourier domains, it is termed as synthetic aperture imaging based on Kramers-Kronig relations (KKSAI). It lies in the intersection of Fourier ptychographic microscopy (FPM), quantitative differential phase contrast (DPC) imaging and digital holographic microscopy. However, it owns several advantages over all these existing imaging modalities. (i) Compared with the aperture scanning FPM \cite{ou2016aperture}, it only needs 2 measurements in principle, thus greatly increasing the imaging speed and reducing the raw data volume. Also, its reconstruction algorithm is iteration-free and parameter-free; (ii) Compared with DPC \cite{lu2016quantitative}, it waives the weak sample assumption, which requires both the absorption and phase modulation magnitude of sample cannot be too large, otherwise will fail DPC. Furthermore, our method does not need to acquire additional amplitude image to reconstruct the complex field; (iii) Compared with in-line holography \cite{zhang2018twin}, its reconstruction is not bothered by twin-image issue, guaranteed by the band-limited signal analyticity. Compared with off-axis holography \cite{baek2019kramers} and its variant \cite{zheng2017digital}, no reference arm is needed and the space-bandwidth product (SBP) can be increased by 3-4 fold under the same detector, by allowing the overlapping of self and cross interference terms.

As we will see, our method bridges all these imaging modalities, which may open a new way of understanding current computational imaging under microscope.
\section{Principle}
\begin{figure}[htbp]
\centering
\includegraphics[width=\linewidth]{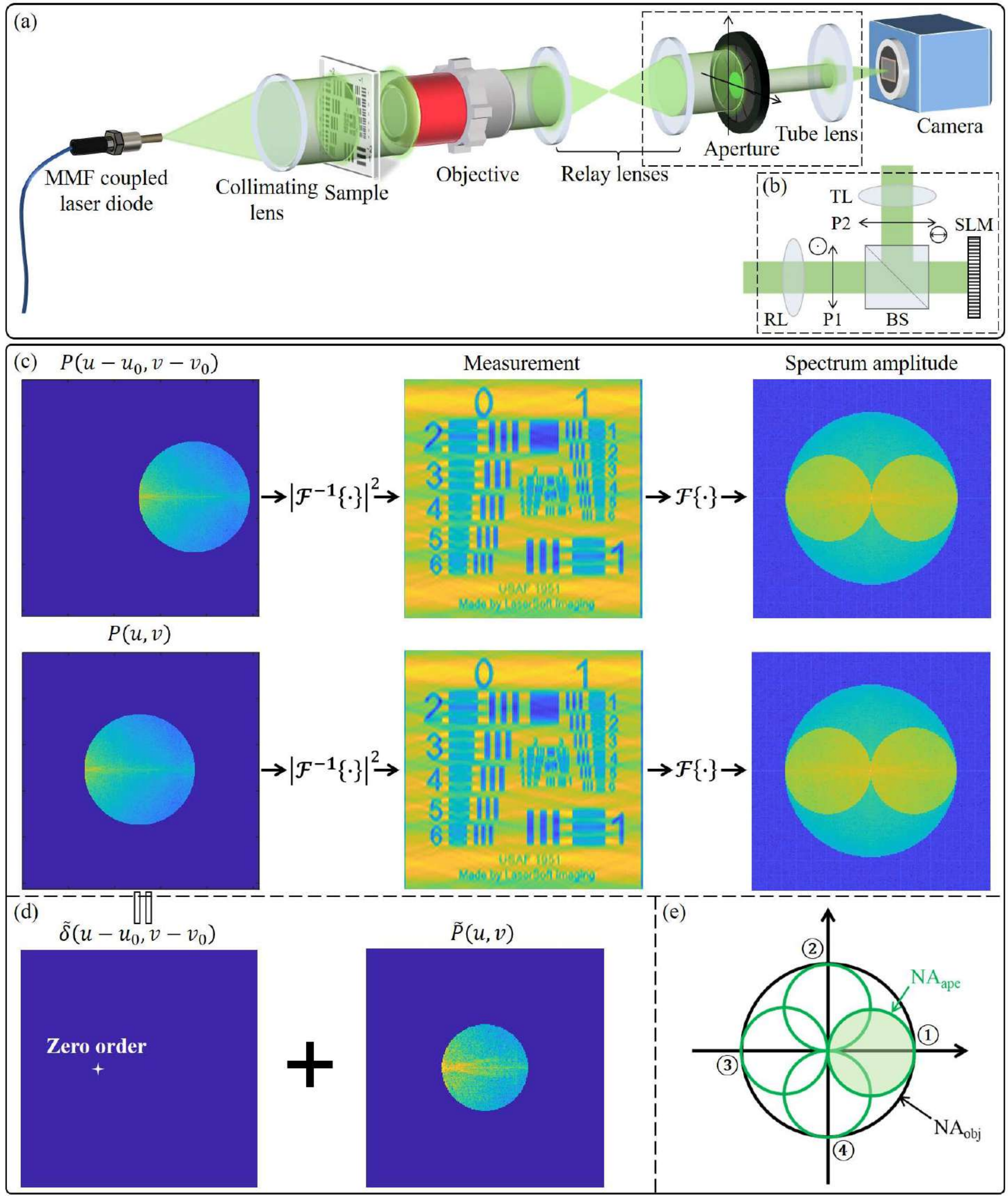}
\caption{Principle of KKSAI. (a) Experimental setup schematic of the proposed method, where the pupil modulation is achieved by an iris diaphragm. Later, to realize more general modulation shape on the pupil plane and achieve the real-time imaging, a SLM-based modulation module shown in (b) replaced the dash-line box in (a). (c) Measurement signal equivalence of two different settings due to the frequency shifting property of Fourier transform. (d) Mathematical decomposition of $P(u,v)$ into a offset delta function $\tilde{\delta}(u-u_0,v-v_0)$ and a hypothetical sample function $\tilde{P}(u,v)$. (e) A simple scanning strategy. MMF, multi-mode fiber; RL, relay lens; P, polarizer; BS, beam splitter; SLM, spatial light modulator; TL, tube lens.
}
\label{fig1}
\end{figure}

As a computational imaging method, sensing process and reconstruction algorithm are co-designed. The imaging system is first introduced. As shown in Fig. \ref{fig1} (a), it is a conventional wide-field microscope, which physically can be simplified as a 4$f$ system. For the convenience of discussion, the coordinate on the sample plane, pupil plane and detector sensor plane is denoted as $(x^\prime, y^\prime)$, $(u, v)$ and $(x, y)$ respectively. In experiment, the pupil plane is relayed outside the objective (10X Mitutoyo Plan Apo infinity corrected objective, 0.28 NA) for easier amplitude modulation, which is realized by an iris diaphragm in the proof-of-concept setup of Fig. \ref{fig1} (a) or a SLM-based module in Fig. \ref{fig1} (b). The latter consists of a reflective mode liquid crystal on silicon (Holoeye LC-R 1080) SLM and a pair of linear polarizers (P1 and P2) with their polarization direction orthogonal to each other in order to maximize the amplitude modulation contrast. The camera (Allied Vision Prosilica GX 6600) pixel size is 5.5 µm. Since our method still belongs to the coherent imaging regime, the illumination is provided by a laser diode (Thorlabs DJ532-40) with the central wavelength of 532 nm, coupled into a multi-mode fiber (Thorlabs FT200emtcustom, 0.39 NA, Ø200 µm). The fiber is vibrated by a motor to wash out the speckle in the output.

During the imaging, the pupil will be sequentially modulated by binary masks, whose edge must exactly cross the pupil center. To clarify the principle, a simple pupil scanning strategy shown in Fig. \ref{fig1} (e) is taken as an example, where the binary mask is a circular aperture. When it scans to the green shade location, the captured image by the square-law detector will be the central top measurement in Fig. \ref{fig1} (c). The frequency band it covers is denoted by $P(u-u_0, v-v_0)$. After Fourier transforming the measurement, a spectrum similar to the one of off-axis hologram is obtained. To explain this observation, two mathematical tools are adopted here. Firstly, due to the frequency shifting property of Fourier transform and the phase loss of square-law detector, the same measurement will be acquired even if the frequency band is shifted to be the pupil plane center, denoted by $P(u, v)$. Then, it is hypothetically separated into a Dirac delta function $\tilde{\delta}(u-u_0,v-v_0)$ and a 'new' sample function $\tilde{P}(u,v)$, shown in Fig. \ref{fig1} (d). Thus, the measurement can be expressed as
\begin{equation}
\begin{aligned}
I &=\left|\mathcal{F}^{-1}\left(\tilde{P}(u, v)+\tilde{\delta}\left(u-u_{0}, v-v_{0}\right)\right)\right|^{2} \\
&=\left|\tilde{p}(x, y)+\exp \left(i u_{0} x, i v_{0} y\right)\right|^{2} \\
&=|\tilde{p}(x, y)|^{2}+1+\tilde{p}^{*}(x, y) \exp \left(i u_{0} x, i v_{0} y\right)\\
&\hspace{3.2mm}+\tilde{p}(x, y) \exp \left(-i u_{0} x,-i v_{0} y\right), 
\end{aligned}
\end{equation}
where $\tilde{p}(x, y)$ is the inverse Fourier transform of $\tilde{P}(u, v)$ and * denotes the the complex conjugate. Then its spectrum is
\begin{equation}
\begin{aligned}
\mathcal{F}\{I\} &=\tilde{P}(u, v) \medwhitestar \tilde{P}(u, v)+\tilde{\delta}(u,v)+\tilde{P}^{*}\left(-\left(u-u_{0}\right),-\left(v-v_{0}\right)\right)\\
&\hspace{3.2mm}+\tilde{P}\left(u+u_{0}, v+v_{0}\right),
\end{aligned}
\end{equation}
where $\medwhitestar$ represents cross-correlation. It can be clearly seen that the first two terms correspond to the self-interference terms in the off-axis hologram and the other two are cross terms.

Usually, it is required in the off-axis hologram that the self- and cross-interference terms be separable in the Fourier domains by adjusting the reference wave incidence angle. Then one cross term will be cropped out and inverse Fourier transformed to obtain the whole complex field. However, a recent study \cite{baek2019kramers} relaxed to it that as long as the cross-interference terms do not overlap each other, the complex field can be reconstructed. Their mathematical framework is summarized as follows. Assuming the sample and reference wave function is $S(\vec{r})$ and $R(\vec{r})$, where $\vec{r}$ is vector representative of $(x,y)$ coordinate, the hologram can be expressed as $I=|S+R|^{2}$.Then,
\begin{equation}
    \frac{I}{|R|^{2}}=\left|\frac{S}{R}+1\right|^{2},
\end{equation}
\begin{equation}
    \ln \left|\frac{S}{R}+1\right|=\frac{1}{2}\left(\ln (I)-\ln \left(|R|^{2}\right)\right).
\end{equation}
Let $X=\ln \left(\frac{S}{R}+1\right)=\operatorname{Re}(X)+j \operatorname{Im}(X)$. We can have
\begin{equation}
    e^{X}=\frac{S}{R}+1=e^{R e(X)} e^{j I m(X)},
\end{equation}
\begin{equation}
    \left|\frac{S}{R}+1\right|=e^{R e(X)},
\end{equation}
\begin{equation}
    \operatorname{Re}(X)=\ln \left|\frac{S}{R}+1\right|=\frac{1}{2}\left(\ln (I)-\ln \left(|R|^{2}\right)\right).
\end{equation}
Considering $R$ is a quasi-plane wave with a wavelength of $\lambda$, it can be expressed as $R(\vec{r})=R_{0} e^{-i 2 \pi \vec{\rho}_{R} \cdot \vec{r}}$, where $\vec{\rho}_R$ is its spatial frequency vector in $(u,v)$ coordinate with its amplitude $|\vec{\rho}_R|=\sin \theta / \lambda$. Thus, its amplitude $R_0$ is ideally a constant and can be calibrated in experiment. Then, $\operatorname{Re}(X)$ can be directly obtained from the measurement $I$ and $R_0$. Now, if $X$ was analytical, KK relations would pin down $\operatorname{Im}(X)$ from $\operatorname{Re}(X)$ and $X$ would be fully determined. Consequently, the sample wave function ${S}$ can be derived from $X$.

Therefore, the problem becomes under what condition $X$ meets analyticity condition. Defining $\alpha=S/R$, we have
\begin{equation}
    X=\log (\alpha+1)=\sum_{n=0}^{\infty} \frac{(-1)^{n}}{n+1} \alpha^{n+1},
\end{equation}
which shows that the analyticity of $X$ depends on the analyticity of $\alpha$. Assuming $R_0=1$,
\begin{equation}
    \alpha(\vec{r})=S(\vec{r}) e^{i 2 \pi \vec{\rho}_{R} \cdot \vec{r}}.
\end{equation}

Now we want to introduce Titchmarsh theorem \cite{titchmarsh1948introduction} to discuss the analyticity of $\alpha$. The theorem states the following conditions for a complex-valued function $f(t)$ that is square integrable over the real $t$-axis are equivalent:
\begin{itemize}
    \item The real and imaginary parts of $f(t)$ are Hilbert transforms of each other.
    \item The Fourier transform $\mathcal{F}(f)(\omega)$ is 0 or vanishes rapidly for $\omega$ < 0.
\end{itemize}

\begin{figure}[htbp]
\centering
\includegraphics[width=\linewidth]{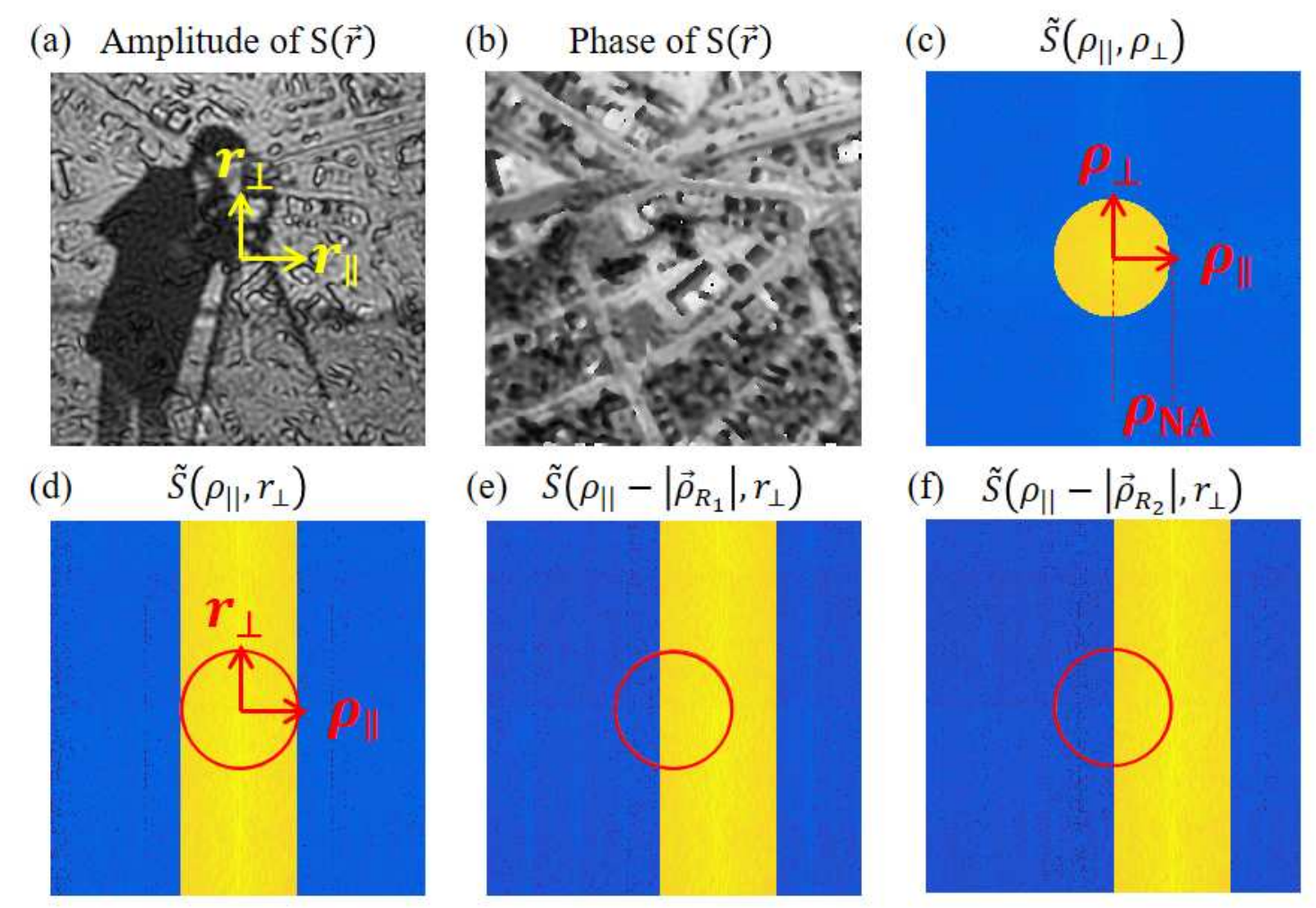}
\caption{Titchmarsh theorem applied to a band-limited signal. (a) Amplitude and (b) phase of simulated band-limited signal with bandwidth of $\rho_\text{NA}$. (c) Logarithm of its 2D Fourier amplitude spectrum. (d) Logarithm of its 1D Fourier amplitude spectrum along $r_{\|}$ axis and its shifted copies by (e) $|\vec{\rho}_{R_1}|<\rho_\text{NA}$ and (f) $\vec{\rho}_{R_2}=\rho_\text{NA}$. 
}
\label{fig2}
\end{figure}

A new axis of $\rho_{\|}=\vec{\rho}_{R}/|\vec{\rho}_{R}|$ is defined on the $(u,v)$ plane with its perpendicular direction $\rho_{\perp}$. Their corresponding axis pair on the $(x,y)$ plane is $r_{\|}$ and $r_{\perp}$. Denoting the 1D Fourier transform of $S(\vec{r})$ along $r_{\|}$ as $\tilde{S}\left(\rho_{\|}, r_{\perp}\right)$, $S(\vec{r})$ can be expressed as
\begin{equation}
    S(\vec{r})=\int_{-\infty}^{\infty} \tilde{S}\left(\rho_{\|}, r_{\perp}\right) e^{i 2 \pi \rho_{\|} \cdot r_{\|}} d \rho_{\|}.
\end{equation}
According to frequency shift property, the 1D Fourier transform of $\alpha(\vec{r})$ along $r_{\|}$ is
\begin{equation}
    \tilde{\alpha}\left(\rho_{\|}, r_{\perp}\right)=\tilde{S}\left(\rho_{\|}-\left|\vec{\rho}_{R}\right|, r_{\perp}\right).
\end{equation}
One important fact is that $S(\vec{r})$ is a band-limited signal because the microscope is a diffraction-limited system characterized by numerical aperture (NA). Thus, $\alpha$ is a complex-valued square integrable function along $r_{\|}$ and Titchmarsh theorem can be effective here. As long as $\tilde{\alpha}\left(\rho_{\|}, r_{\perp}\right)=0$ for $\rho_{\|}<0$, its real and imaginary part are Hilbert transforms of each other, showing its analyticity.

To make the deduction above clear, a simulated $S(\vec{r})$ is shown in Fig. \ref{fig2} (a-c). Its 1D Fourier transform along $r_{\|}$ axis in Fig. \ref{fig2} (d) cannot guarantee that $\tilde{\alpha}\left(\rho_{\|}, r_{\perp}\right)=0$ for $\rho_{\|}<0$. But when it is shifted by $|\vec{\rho}_{R}|$ until $|\vec{\rho}_{R}|=\rho_\text{NA}$ like the case in Fig. \ref{fig2} (f), this condition is met and $\alpha$ is analytical.

To sum up, when $|\vec{\rho}_{R}| \geq \rho_\text{NA}$, X meets the analyticity condition whereby KK relations can reconstruct $\operatorname{Im}(X)$ from $\operatorname{Re}(X)$ and the complex field will be known. The critical situation $|\vec{\rho}_{R}|=\rho_\text{NA}$ means that the two cross-interference terms in Fourier domains are tangent to each other exactly, which is just the case in Fig. \ref{fig1} (c). In other words, as long as the amplitude modulation mask edge cross the spectrum origin and do not overlap its mirrored shape, the hypothetical reference wave in the measurement will satisfy the critical situation to apply KK relations.

\begin{figure}[htbp]
\centering
\includegraphics[width=\linewidth]{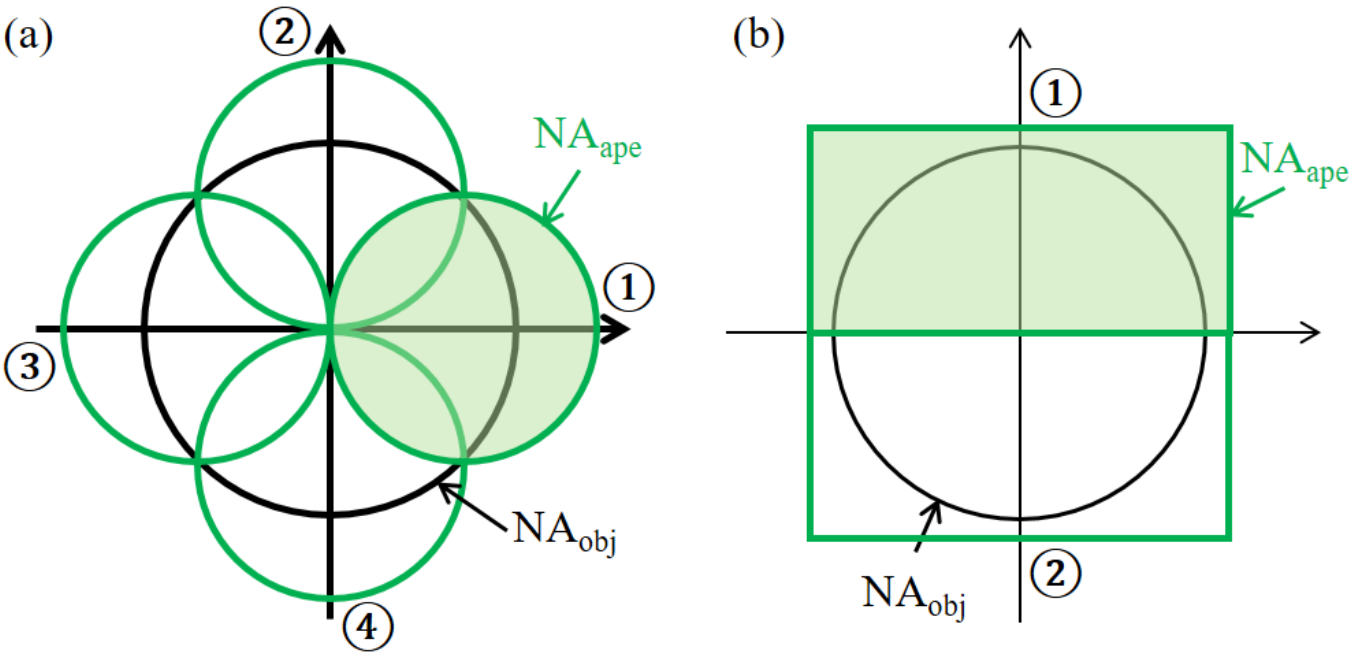}
\caption{Scanning scheme examples to cover the entire pupil. (a) Four circular apertures; (b) Two rectangular apertures. \textcircled{\small{1}} $\cdots$ \textcircled{\small{4}} label the sequential measurements.
}
\label{fig3}
\end{figure}

According to this conclusion, our computational imaging method is designed as follows. Following the aforementioned conditions, at least 2 apertures are required to cover the whole pupil of imaging system, shown in Fig. \ref{fig3} (b). In the proof-of-concept system, an iris diaphragm is utilized so another viable but not the most efficient scanning scheme is displayed in Fig. \ref{fig3} (a). All the scanning apertures keep their edge cross the pupil origin exactly. If not stated, Figure \ref{fig3} (a) is our default scanning scheme in order to compare our method with other existing ones. The aperture function is denoted as $A(\vec{\rho})$. The sequential measurements are $I_{i}$ ($i=1,\cdots,N$) with aperture offset $\vec{\rho}_i$ ($ |\vec{\rho}_i| = \rho_{NA}$) on the pupil plane.

Another crucial point when applying our proposed method is that the camera sampling rate must satisfy the Nyquist limit $2 \rho_{NA}$ to avoid the sub-sampling aliasing.

The reconstruction steps are illustrated in algorithm \ref{alg1}. The for-loop part completes phase recovery based on KK relations, which can be summarized into four steps. (i) Generate virtual reference wave $R$; (ii) Calculate Hilbert kernel according to the aperture offset; (iii) Reconstruct $\operatorname{Im}(X)$ with KK relations by directional Hilbert transform; (iv) Recover the complex field corresponding to each frequency band. Finally the frequency bands are stitched together using image blending method in Fourier domains. Here the simplest way, alpha blending, is adopted where $\epsilon$ is for numerical stability. The output $P(\vec{\rho})$ contains all the information of complex wave-field which can be detected by any diffraction-limited imaging system. For experimental datasets, the aperture function $A(\vec{\rho})$ and its offsets can be post-calibrated to sub-pixel accuracy by \cite{eckert2018efficient}.

\begin{algorithm}
\caption{KKSAI reconstruction algorithm}\label{alg1}
\begin{algorithmic}[1]
\State Input: $I_i, A(\vec{\rho}), \vec{\rho}_i$
\ForAll{$i$}\Comment{Phase recovery for each aperture}
\State $R=e^{-i 2 \pi (-\vec{\rho}_i) \cdot \vec{r}}$
\State $\operatorname{Re}(X)=\frac{1}{2}\left(\ln (I_i)-\ln \left(|R|^{2}\right)\right)$
\State $H = \operatorname{sgn}\left(\vec{\rho} \cdot \vec{\rho}_i\right)$
\State $\operatorname{Im}(X)=-i\mathcal{F}^{-1}\left\{\mathcal{F}\{\operatorname{Re}(X)\}(\vec{\rho}) \times H\right\}$
\State $X=\operatorname{Re}(X)+i\operatorname{Im}(X)$
\State $S=\left(\exp{(X)}-1\right) \times R$
\State $Band(\vec{\rho})=\mathcal{F}\{S\} \times A(\vec{\rho})$
\EndFor
\State $P(\vec{\rho})=\sum_{i=1}^{N}Band(\vec{\rho}-\vec{\rho}_i)/\left(\sum_{i=1}^{N}A(\vec{\rho}-\vec{\rho}_i)+\epsilon\right)$
\State Output: $P(\vec{\rho})$
\end{algorithmic}
\end{algorithm}

\section{Simulation}
In this section, a series of simulations are conducted to verify the performance of our proposed method and compare it with two existing imaging modalities, DPC \cite{lu2016quantitative} and FPM \cite{sun2018high}. To be fair, the same dataset $I_i$ is used in all the methods.

\begin{figure}[htbp]
\centering
\includegraphics[width=\linewidth]{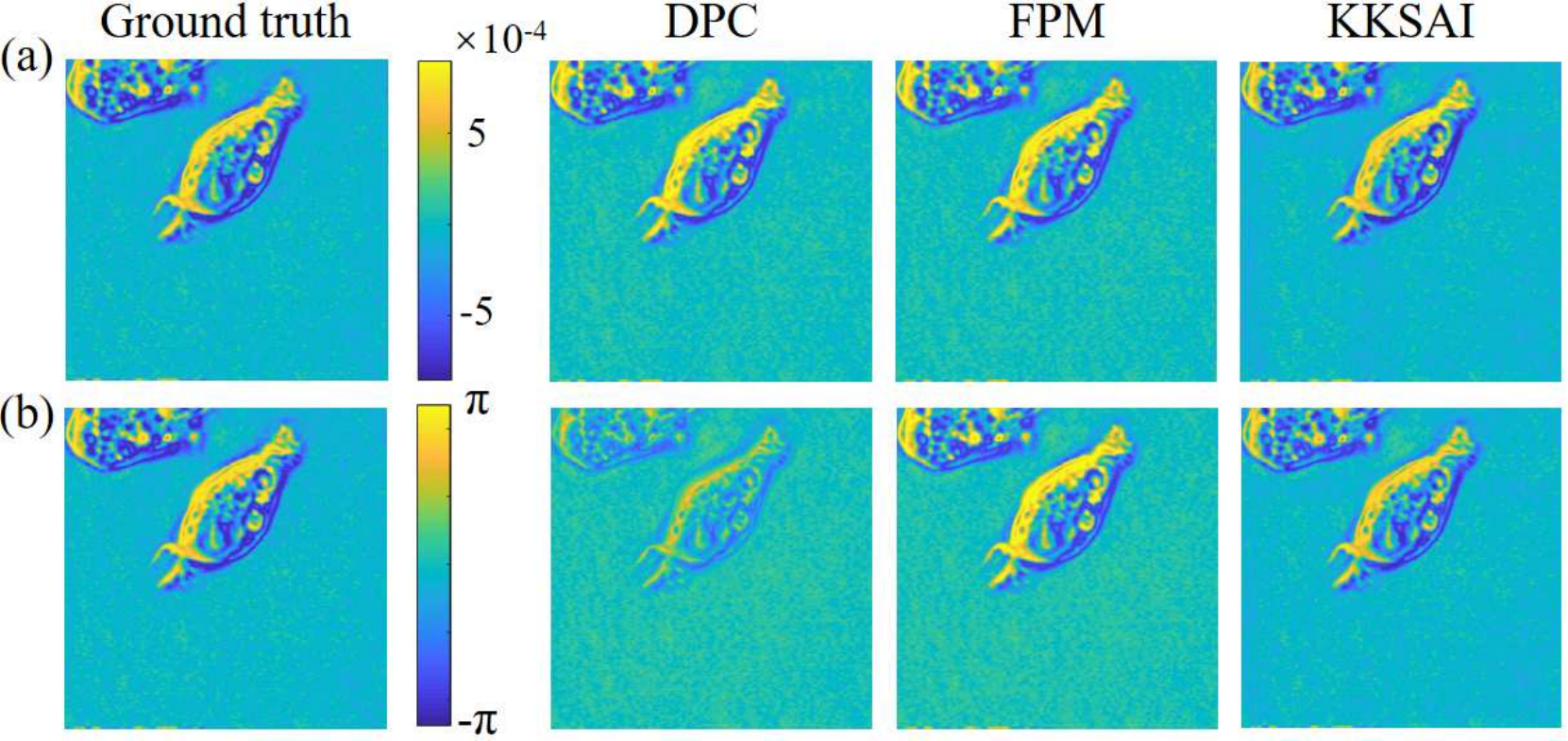}
\caption{Reconstructions of phase-only sample by two existing imaging modalities and our method. (a) Weak phase sample; (b) Strong phase sample.
}
\label{fig4}
\end{figure}

Firstly, since phase recovery is key to our problem, pure phase samples with small and large phase difference are simulated and studied, as shown in Fig. \ref{fig4}. From the visual evaluation on the reconstructions, KKSAI obtains the phase map closest to ground truth whatever the phase magnitude is. To quantitatively compare their performances, mean square error (MSE) and feature similarity (FSIM) index \cite{zhang2011fsim} between reconstruction and ground truth are calculated in Tab. \ref{tab1}.

\begin{table}[htbp]
\centering
\caption{\bf Quantitative evaluation of reconstructions in Fig. \ref{fig4}}
\begin{tabular}{c|c|c|c|c}
\hline
\multicolumn{2}{c|}{Metric} & DPC & FPM & KKSAI \\
\hline
\multirow{3}{*}{(a)} & MSE & $4.80 \times 10^{-9}$ & $4.53 \times 10^{-9}$ & $7.84 \times 10^{-10}$\\
& FSIM & 0.9999 & 1.0000 & 0.9998 \\
& Time(s) & 2.01 & 28.43 & 2.12\\
\hline
\multirow{3}{*}{(b)} & MSE & 0.1711 & 0.0640 & 0.0136\\
& FSIM & 0.9894 & 1.0000 & 0.9973 \\
& Time(s) & 1.67 & 77.33 & 2.03\\
\hline
\end{tabular}\label{tab1}
\end{table}

The results show that FPM's phase reconstructions hold perfect similarity but the absolute error is dramatically larger than other two methods, which means its result has significant constant phase shift. Meanwhile, its iteration framework is quite time-consuming. In MATLAB R2018b on a computer (i7-7700k) with 64GB RAM, its running time is one order of magnitude higher than other two iteration-free methods. 

As for DPC, the weak sample assumption allows applying first-order Taylor expansion approximation and is the foundation of its reconstruction algorithm. When the phase magnitude increases, DPC's phase recovery deviates from the ground truth quickly, whether from MSE or FSIM. This is in accordance with conclusions in \cite{lu2016quantitative}.

\begin{figure}[htbp]
\centering
\includegraphics[width=\linewidth]{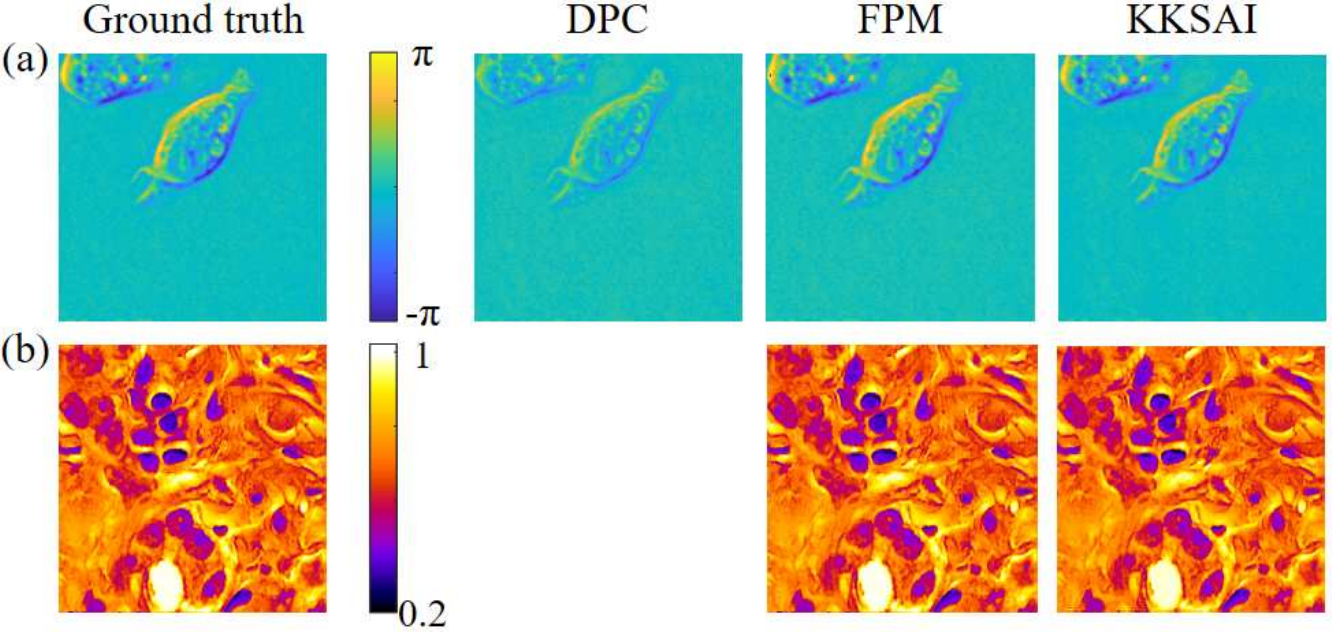}
\caption{Reconstructions of complex-valued sample by two existing imaging modalities and our method. (a) Phase; (b) Amplitude.
}
\label{fig5}
\end{figure}

Next, a complex-valued sample is simulated and its amplitude and phase are displayed in Fig. \ref{fig5} (a) and (b). The quantitative analysis is summarized in Tab. \ref{tab2}. DPC is incapable of reconstructing amplitude without additional measurement. Moreover, the large phase magnitude hampers its prerequisite, resulting in a low recovery accuracy. Judging from the naked eyes and metrics, FPM achieves almost the same satisfactory reconstructions with KKSAI. However, it needs good initialization and parameter settings. Its iteration process is around 40 times longer than KKSAI.

\begin{table}[htbp]
\centering
\caption{\bf Quantitative evaluation of reconstructions in Fig. \ref{fig5}}
\begin{tabular}{c|c|c|c|c}
\hline
\multicolumn{2}{c|}{Metric} & DPC & FPM & KKSAI \\
\hline
\multirow{2}{*}{Phase} & MSE & 0.0531 & 0.0120 & 0.0037\\
& FSIM & 0.9934 & 0.9997 & 0.9976 \\
\hline
\multirow{2}{*}{Amplitude} & MSE & / & $1.55 \times 10^{-8}$ & $3.78 \times 10^{-4}$\\
& FSIM & / & 1.0000 & 0.9965 \\
\hline
\multicolumn{2}{c|}{Time(s)} & 3.83 & 109.37 & 2.82\\
\hline
\end{tabular}\label{tab2}
\end{table}

To sum up, KKSAI outperforms DPC and FPM taking both reconstruction accuracy and computational load into account.

\begin{figure}[htbp]
\centering
\includegraphics[width=\linewidth]{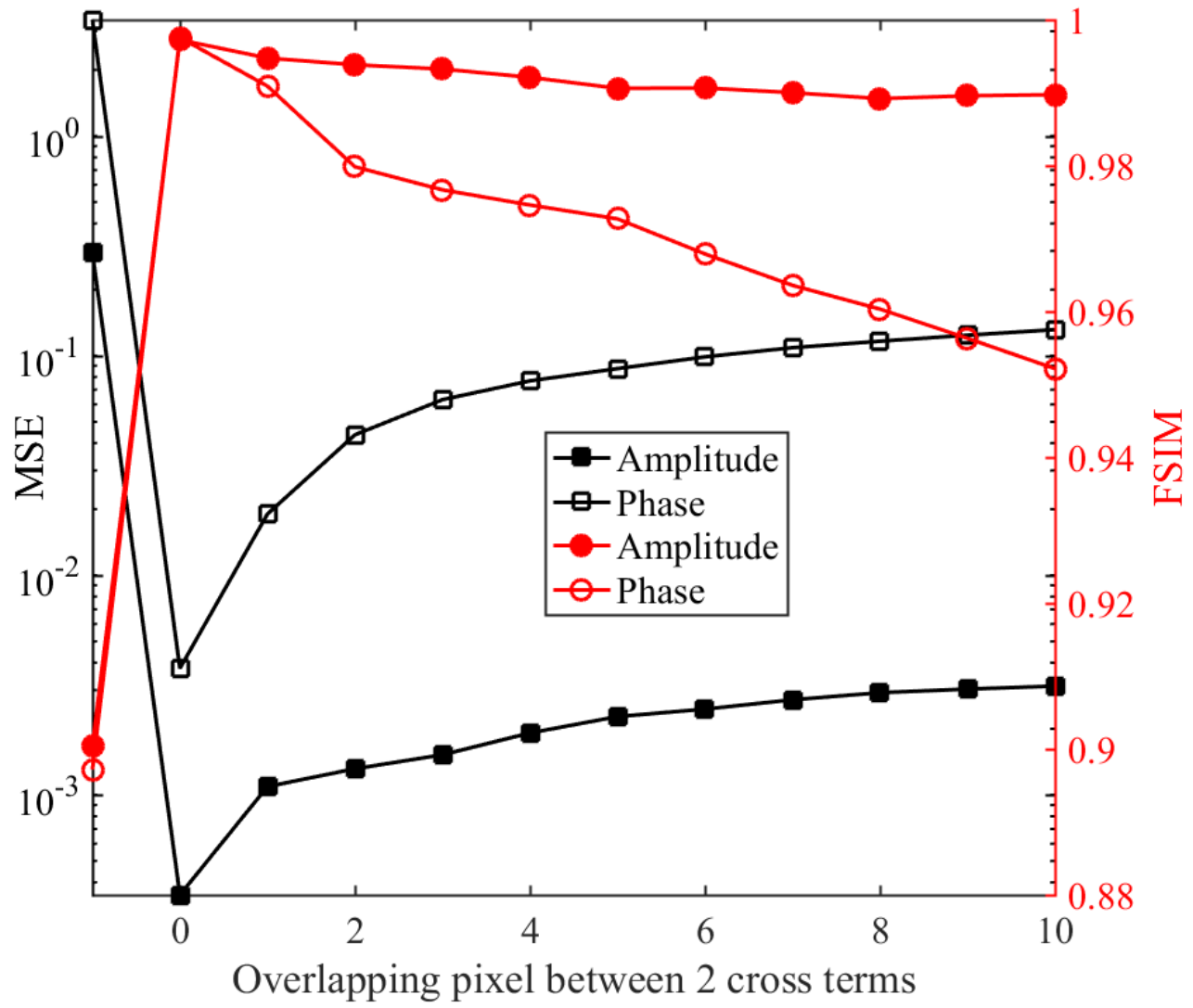}
\caption{The effect of distance between pupil center and aperture edge on the final reconstruction accuracy.
}
\label{fig6}
\end{figure}

Next, an important experimental parameter is simulated and discussed because it will affect the final reconstruction quality. As stated before, the aperture edge needs to exactly cross the pupil center. If it contains the pupil center, the cross-interference terms will overlap, which breaks the analyticity condition. If it excludes the pupil center, the strong DC term cannot be collected and the hologram-like spectrum will not appear in measurements. In simulation, the overlapping between two cross terms is controlled and characterized by pixel number. Figure \ref{fig6} displays how overlapping ruins the reconstruction accuracy. The negative overlapping represents that the pupil center is outside of aperture. The trend of all curves indicates the significance of pupil center just lying on the aperture edge.

\begin{figure}[htbp]
\centering
\includegraphics[width=\linewidth]{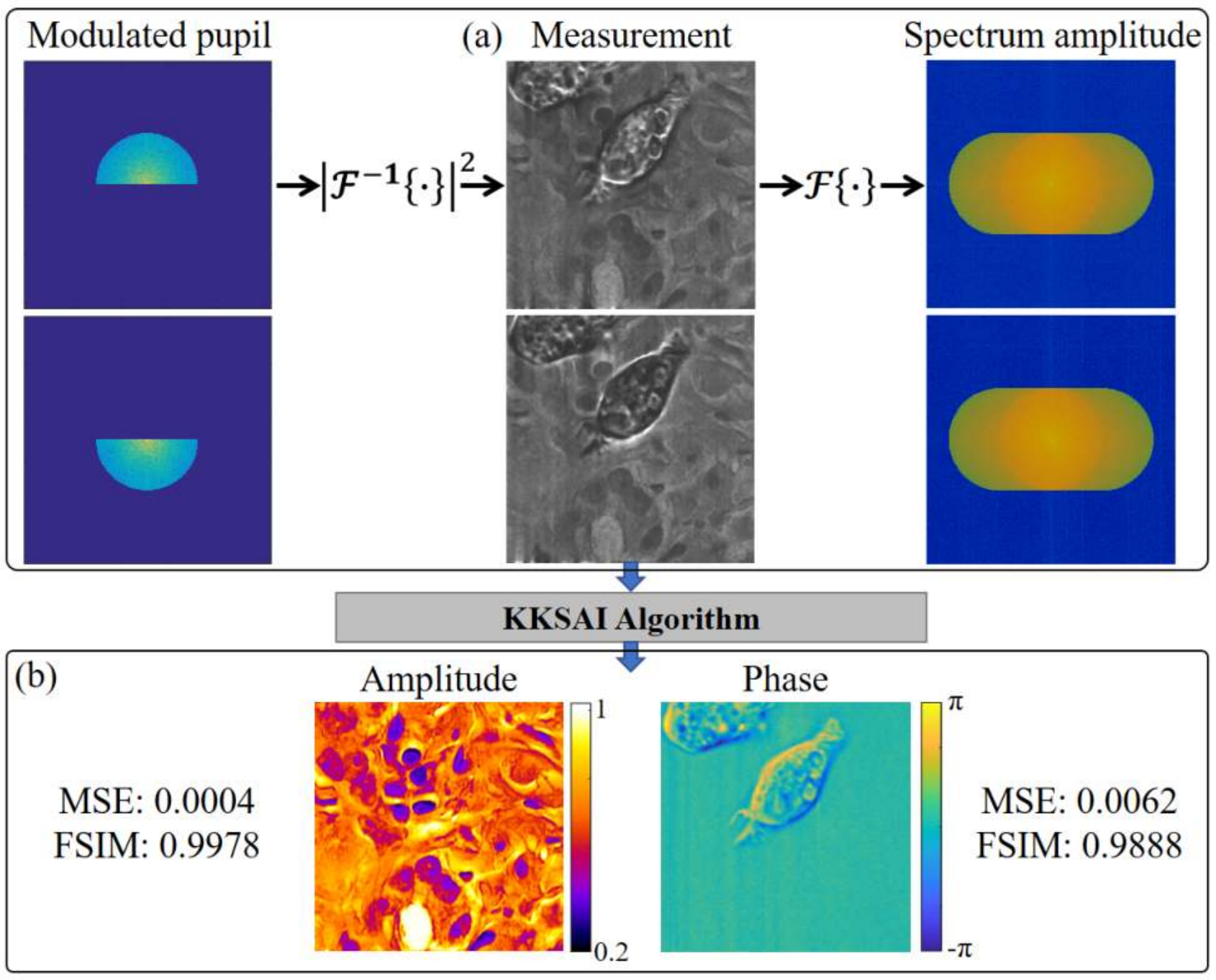}
\caption{KKSAI based on the scanning scheme with the minimal measurements. (a) Sensing; (b) Reconstruction.
}
\label{fig7}
\end{figure}

Finally, the viability of scanning scheme in Fig. \ref{fig3} (b) is verified in simulation. As shown in Fig. \ref{fig7} (a), only a half of pupil passes through in each measurement and there still exists the self and cross terms in their spectrum, although it is a little bit hard to distinguish two cross terms considering they illusorily form a whole circle. By the KKSAI algorithm, the reconstructions in Fig. \ref{fig7} (b) fit very well with the ground truth in Fig. \ref{fig5}.

\section{Experiment}

First, the proof-of-concept experiments are conducted on the setup shown in Fig. \ref{fig1} (a) and the scanning strategy follows Fig. \ref{fig1} (e). A typical plant cell slide sample (AmScope PS100A) is used and it is mathematically complex-valued. 

\begin{figure}[htbp]
\centering
\includegraphics[width=\linewidth]{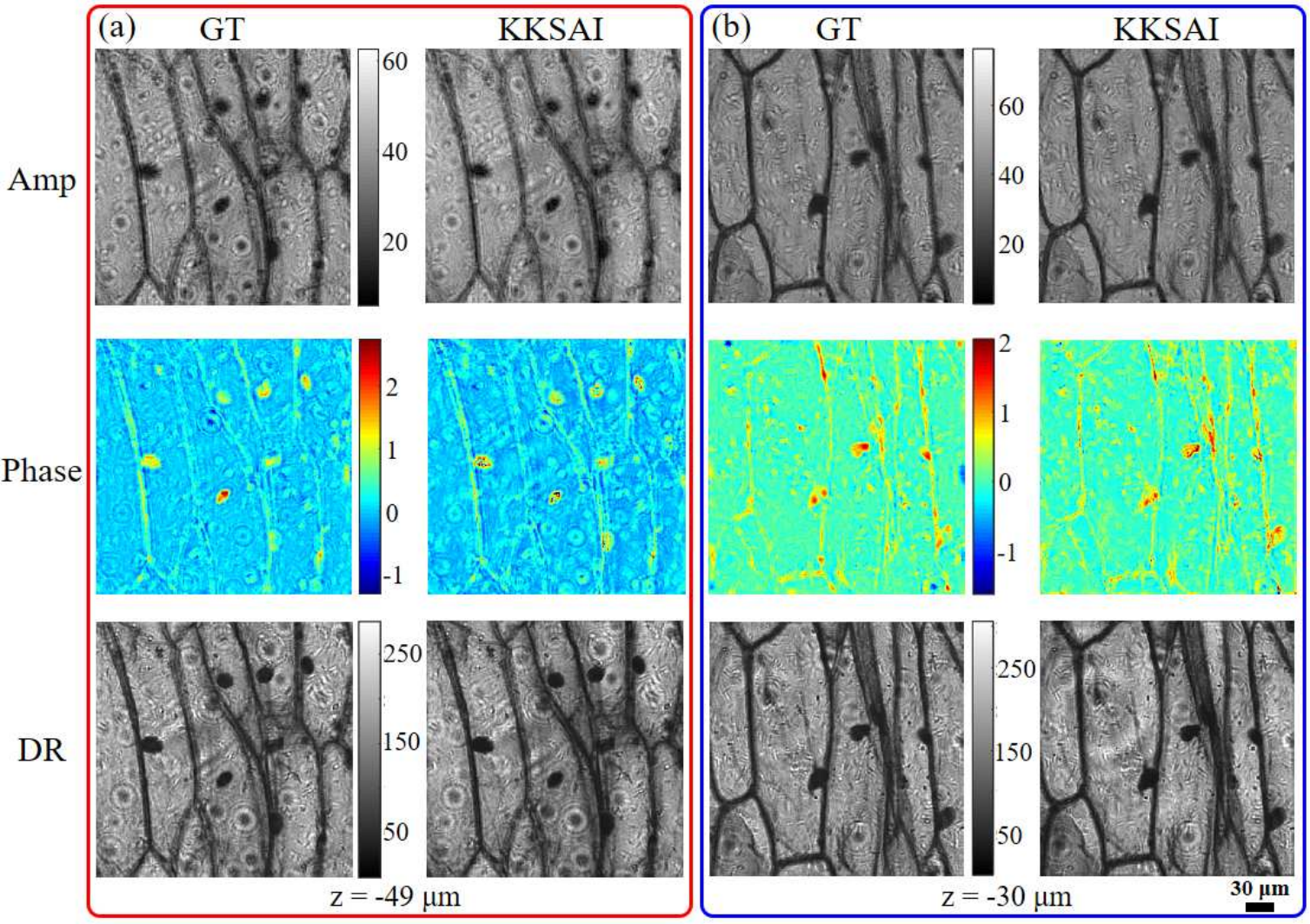}
\caption{Comparison between ground truth (GT) and reconstructions by KKSAI for plant cells sample. (a) ROI1; (b) ROI2. The defocusing distance is labeled at the bottom. Amp, amplitude; DR, digitally refocused amplitude.
}
\label{fig9}
\end{figure}

\begin{figure*}[htbp]
\centering
\includegraphics[width=\linewidth]{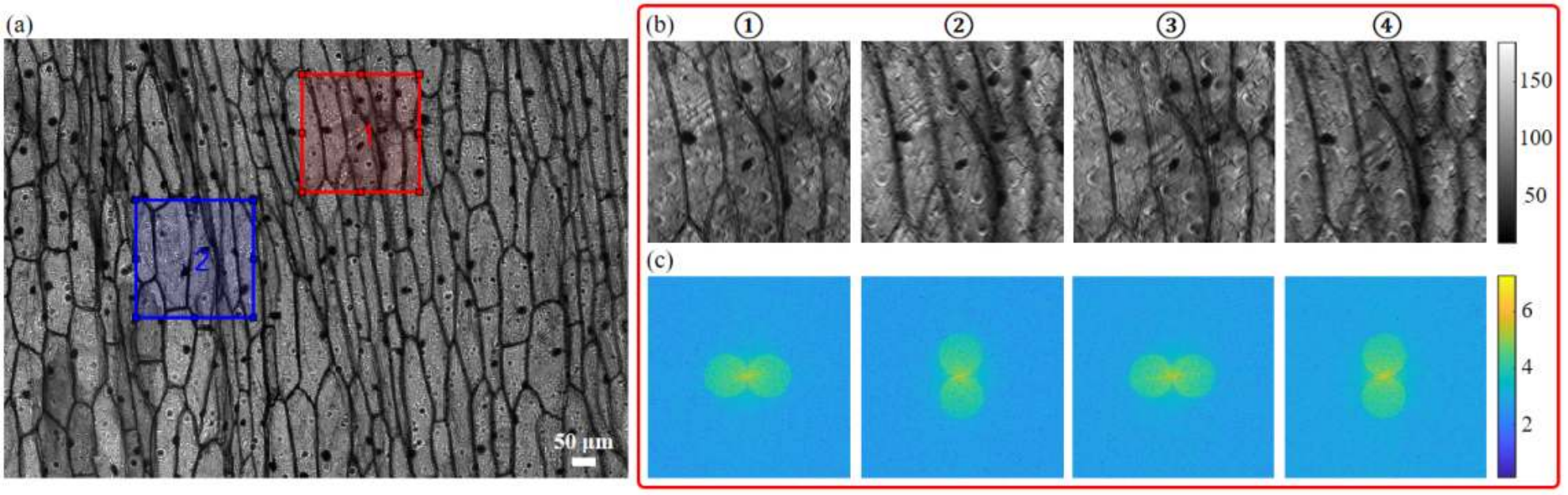}
\caption{Experiment on typical plant cells microslide. (a) Wide-field image of sample with the colorful inset boxes indicating ROIs; (b) KKSAI measurements of ROI1 and (c) the logarithm amplitude spectrum of the measurements. 
}
\label{fig8}
\end{figure*}

Figure \ref{fig8} displays the raw images in the experiment on the plant cells microslide. Obviously shown in Fig. \ref{fig8} (c), their spectrum expresses two similar cross-interference terms in an off-axis hologram, just like Fig. \ref{fig1} (c). Thanks to the delicate alignment of aperture, these two terms are tangent to each other. Feeding measurements into the Algorithm \ref{alg1}, the complex-field reconstructions of two regions of interest (ROIs) are obtained and presented in Fig. \ref{fig9}.

Since the absolute ground truth in experiment is hard to acquire unlike in simulation, an individual FPM experiment was performed on the same sample. 47 subaperture images with an overlapping rate of about 85\% are stitched into a high-resolution reconstruction by FPM algorithm, which is taken as the golden rule here. Evaluated with naked eyes, both the amplitude and phase recovery of KKSAI are quite close to the ground truth. To further assess the phase retrieval ability of KKSAI, digital refocusing by the angular spectrum method \cite{goodman2005introduction} are applied to it. With the Gini of the gradient index \cite{zhang2017edge}, the best focusing distance from KKSAI reconstruction is -49 µm and -30 µm for two ROIs, the same as the ground truth. 

\begin{table}[htbp]
\centering
\caption{\bf Quantitative evaluation of KKSAI reconstructions in Fig. \ref{fig9}}
\begin{tabular}{c|c|c|c|c}
\hline
\multicolumn{2}{c|}{Metric} & Amp & Phase & DR \\
\hline
\multirow{2}{*}{ROI1} & NCC & 0.9410 & 0.8052 & 0.9258 \\
& FSIM & 0.9925 & 0.9769 & 0.9893 \\
\hline
\multirow{2}{*}{ROI2} & NCC & 0.9341 & 0.7832 & 0.9151 \\
& FSIM & 0.9914 & 0.9727 & 0.9884 \\
\hline
\end{tabular}\label{tab3}
\end{table}

Quantitative evaluation is seen in Tab. \ref{tab3}. Since the ground truth utilizes more images than KKSAI, there exists an energy mismatch between them. Thereby MSE is not a reasonable metric anymore but instead normalized cross correlation (NCC) is used to assess the global error. It is clear that both KKSAI reconstruction and its digitally refocused version match the ground truth very well, with a high NCC or FSIM value. 

\begin{figure}[htbp]
\centering
\includegraphics[width=\linewidth]{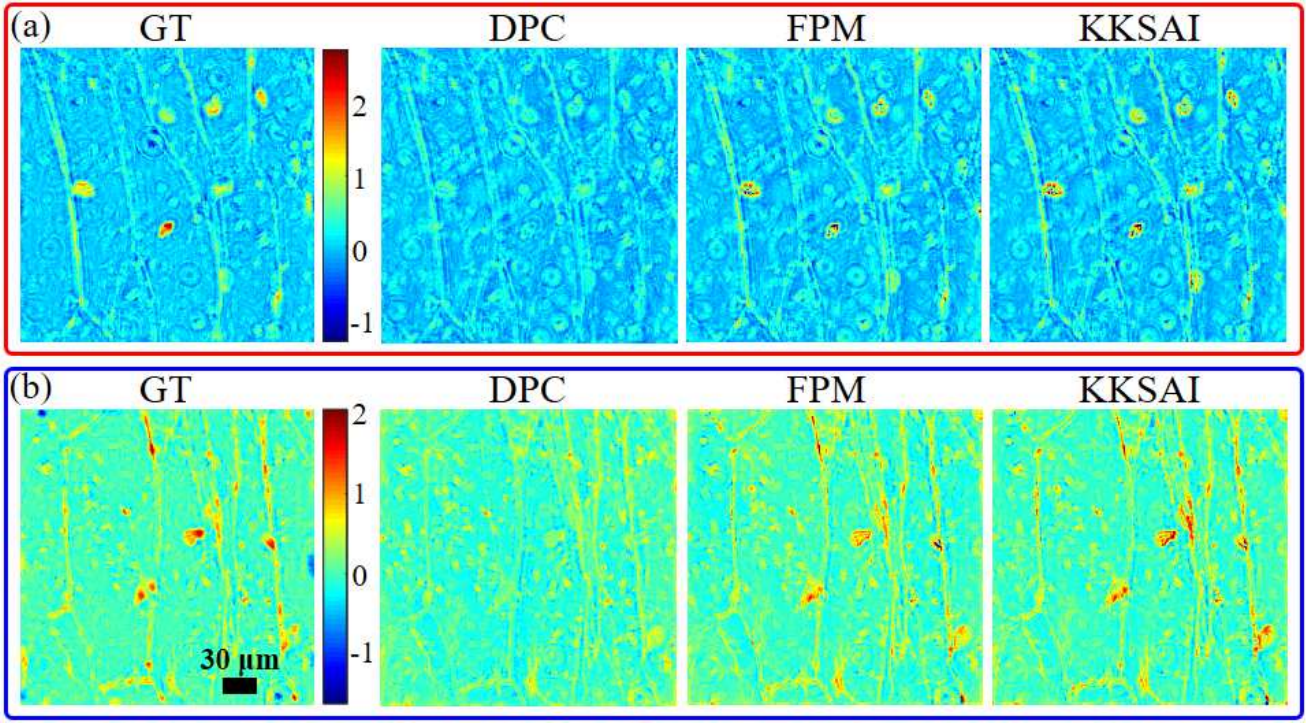}
\caption{Comparison of phase recovery from two existing methods and KKSAI for plant cells sample. (a) ROI1; (b) ROI2.}
\label{fig10}
\end{figure}

Furthermore, the phase recovery from two existing methods and our KKSAI is compared. Fed with the same measurements, their reconstruction results are summarized in Fig. \ref{fig10} and Tab. \ref{tab4}. Since the weak sample assumption is unsatisfied, DPC has a poor reconstruction, which can be easily told from Fig. \ref{fig10}. As for FPM, the low overlapping rate of scanning apertures coupled with experimental noises makes it perform not as well as in simulation. From both NCC and FSIM metric, KKSAI phase reconstruction is better than other two's.

\begin{table}[htbp]
\centering
\caption{\bf Quantitative evaluation of KKSAI reconstructions in Fig. \ref{fig10}}
\begin{tabular}{c|c|c|c|c}
\hline
\multicolumn{2}{c|}{Metric} & DPC & FPM & KKSAI \\
\hline
\multirow{2}{*}{ROI1} & NCC & 0.7198 & 0.7955 & 0.8052 \\
& FSIM & 0.9613 & 0.9749 & 0.9769 \\
\hline
\multirow{2}{*}{ROI2} & NCC & 0.6801 & 0.7652 & 0.7832 \\
& FSIM & 0.9575 & 0.9712 & 0.9727 \\
\hline
\end{tabular}\label{tab4}
\end{table}

[Pure-phase sample \& real-time imaging experiment by SLM]

\section{Conclusion}

In this paper, a new computational method to reconstruct the whole complex light field is reported. Based on observation of the measurement spectrum from amplitude modulated pupil, we found the analogy between it and the off-axis hologram. Then, thanks to the analyticity of band-limited signal under all diffraction-limited imaging systems, a specially designed quantity directly related to the complex light field can be fully determined through KK relations. As a computational imaging method, the proposed KKSAI co-designed the sensing part and the reconstruction algorithm. From the perspective of sensing, it requires much less measurements than modalities like FPM. From the perspective of reconstruction algorithm, it is iteration-free and does not need any knowledge of sample priors, like DPC.

In spite of the mathematical analogy to off-axis hologram, it is distinct from off-axis holographic microscopy due to the absence of real reference wave arm. Moreover, there exists a subtle difference between these two settings. It will benefit the understanding of off-axis holography if clarified.
Mathematically, the off-axis holography can be generalized to be seen as the addition of a band-limited sample spectrum and a delta function in frequency domains. Their relative offset will decide the distance of cross terms. In both KKSAI and off-axis holography \cite{baek2019kramers}, the delta function lies on the edge of sample spectrum. But the difference is that for KKSAI there is a total offset of sample spectrum and delta function such that the delta function lies on the origin and becomes the DC value. This explains a common question coming with our observation. Why are there no interference fringes in the KKSAI measurements? From this point of view, the recovery of any coherent signal with a strong DC peak could benefit from the KK relations if carefully co-designing the sensing and reconstruction algorithm.

It is also worth mentioning that our work here is discussed under the aperture scanning mode but it can be easily transferred to its reciprocal mode, by tilted illumination. The multiplication with a phase factor in spatial domain brought by tilted illumination is equivalent to the offset modulation in spatial frequency domain. Since the mathematical nature stays the same, we did not expand on that in this paper. Another interesting point is that under tilted illumination mode, our method can be realized by lighting up LED elements located on a ring with the illumination NA matching the objective NA. This is known as annular illumination scheme and it has been shown that only under this condition low-frequency phase components can be transferred into intensity \cite{sun2018high}.
Although the conclusion is derived based on the weak sample assumption, the coincidence between this and our findings indicates that keeping the DC component at the pupil/aperture edge is a special sweet point for computational imaging. This may provide a new viewpoint of designing new computational imaging methods.

\bibliography{KKSAI}

\bibliographyfullrefs{KKSAI}


\ifthenelse{\equal{\journalref}{aop}}{%
\section*{Author Biographies}
\begingroup
\setlength\intextsep{0pt}
\begin{minipage}[t][6.3cm][t]{1.0\textwidth} 
  \begin{wrapfigure}{L}{0.25\textwidth}
    \includegraphics[width=0.25\textwidth]{john_smith.eps}
  \end{wrapfigure}
  \noindent
  {\bfseries John Smith} received his BSc (Mathematics) in 2000 from The University of Maryland. His research interests include lasers and optics.
\end{minipage}
\begin{minipage}{1.0\textwidth}
  \begin{wrapfigure}{L}{0.25\textwidth}
    \includegraphics[width=0.25\textwidth]{alice_smith.eps}
  \end{wrapfigure}
  \noindent
  {\bfseries Alice Smith} also received her BSc (Mathematics) in 2000 from The University of Maryland. Her research interests also include lasers and optics.
\end{minipage}
\endgroup
}{}

\end{document}